\documentclass[aps,prb,reprint,10pt,fleqn,showpacs,showkeys,floatfix,superscriptaddress]{revtex4-1}
\usepackage{graphicx}
\usepackage{amssymb}
\usepackage{amsmath}
\usepackage{bm}
\usepackage{dcolumn}
\usepackage{graphicx}
\usepackage{dcolumn}
\usepackage{bm}

\begin{document}

\title{Pressure-Induced Enhancement of the Magnetic Anisotropy in Mn(N(CN)$_{2}$)$_{2}$}

\author{P. A. Quintero}
\affiliation{Department of Physics and National High Magnetic Field Laboratory, University of Florida, Gainesville, FL 32611-8440, USA}
\author{D. Rajan}
\affiliation{Department of Chemistry, University of Florida, Gainesville, FL 32611-7200, USA}
\author{M. K. Peprah}
\affiliation{Department of Physics and National High Magnetic Field Laboratory, University of Florida, Gainesville, FL 32611-8440, USA}
\author{T. V. Brinzari}
\affiliation{Department of Physics and National High Magnetic Field Laboratory, University of Florida, Gainesville, FL 32611-8440, USA}
\author{R. S. Fishman}
\affiliation{Materials Science and Technology Division, Oak Ridge National Laboratory, Oak Ridge, TN 37831-6190, USA}
\author{D. R. Talham}
\affiliation{Department of Chemistry, University of Florida, Gainesville, FL 32611-7200, USA}
\author{M. W. Meisel}
\affiliation{Department of Physics and National High Magnetic Field Laboratory, University of Florida, Gainesville, FL 32611-8440, USA} 

\date{\today}

\begin{abstract}
Using dc and ac magnetometry, the pressure dependence of the magnetization of the three-dimensional antiferromagnetic coordination 
polymer Mn(N(CN)$_{2}$)$_{2}$ was studied up to 12~kbar and down to 8~K. The magnetic transition temperature, $T_c$, increases 
dramatically with applied pressure~$(P)$, where a change from $T_c(P=\text{ambient}) = 16.0$~K to $T_c(P=12.1$~kbar$) = 23.5$~K 
was observed.  In addition, a marked difference in the magnetic behavior is observed above and below 7.1~kbar. Specifically, for 
$P<7.1$~kbar,  the differences between the field-cooled and zero-field-cooled \mbox{(fc-zfc)} magnetizations, the coercive field, 
and the remanent magnetization decrease with increasing pressure. However, for $P>7.1$~kbar, the behavior is inverted. Additionally, 
for $P>8.6$~kbar, minor hysteresis loops are observed. All of these effects are evidence of the increase of the superexchange 
interaction and the appearance of an enhanced exchange anisotropy with applied pressure.

\end{abstract}

\pacs{Valid PACS appear here}

\maketitle

\section{\label{sec:level1}Introduction}

There is a growing interest in the magnetic properties of molecule-based magnets under hydrostatic pressure. Due to the 
compressibility of the compounds containing organic ligands these materials can show enhanced transition temperatures and 
new magnetic behaviors when subject to applied pressure.\cite{DaSilva2013,Miller2014} One interesting example is the compound  
Mn(N(CN)$_{2}$)$_{2}$, which belongs to the isostructural family M(N(CN)$_{2}$)$_{2}$ (M~=~Mn, Fe, Co, Ni). These materials 
have a three-dimensional (3D) rutile-like structure with the metal centers connected by dicyanamide ligands 
\mbox{(N$\equiv$C-N-C$\equiv$N)$^-$}, so each metal is surrounded by a N$_6$ octahedron, and images of the crystal 
structure are readily available in the literature.\cite{Kurmoo1998,Kmety2000}  All the members in the family show long range magnetic 
order attributed to interactions between the metal centers along the \mbox{M-[N-C-N]-M} superexchange path. 
In spite of the similarity of the crystal structures, different metals show strikingly different magnetic behaviors, where 
the Mn and Fe analogues are long-range canted-antiferromagnets while the Co and Ni systems are 
ferromagnets.\cite{Kurmoo1998,Kurmoo1999,Batten1999,Review-Batten2003103} Based on crystallographic information, it has 
been suggested the nature of the magnetic interaction between the metal centers depends solely on the angle between 
the metals and the carbon along the superexchange path $\widehat{\text{M-C-M}}$, where a crossover from non-collinear 
antiferromagnetism to ferromagnetism occurs for a an angle of 142$^{\circ}$.\cite{Kmety2000} However, magnetic measurements 
with compounds of mixed metals and computational studies suggest some other factors beyond such an angle also play a role in 
determining the sign of the superexchange interaction.\cite{Lappas2003,PhysRevB.69.205105}

Previously, dc and ac magnetometry, \cite{Review-Batten2003103} muon-spin rotation, \cite{muons} specific heat, and 
powder neutron diffraction measurements were employed to explore the magnetism of  Mn(N(CN)$_{2}$)$_{2}$.\cite{Kmety2000,Lappas2003} 
Long-range canted-antiferromagnetic ordering is observed below $T_c \approx 16$~K, with the spins of the Mn centers 
oriented in the \textit{ab} crystallographic plane so no component is in the \textit{c} axis and a small 
uncompensated moment is along the \textit{b} axis. In the \textit{ab} plane, the spins show antiparallel arrangement 
along the \textit{a} axis and parallel orientations along the \textit{b} axis.\cite{Kmety2000,Lappas2003} The spin 
canting has been attributed to the Dzyaloshinskii-Moriya (DM) antisymmetric interaction, which also explains the 
magnitude of the canting angle.\cite{Lappas2003}

The M = Fe, Co, and Ni compounds have been previously studied using low field ac magnetometry under pressure.  
The Fe(N(CN)$_{2}$)$_{2}$ and Ni(N(CN)$_{2}$)$_{2}$ compounds show an increase of the transition temperature 
of 26\% and 6\%, respectively, for pressures as large as 17~kbar, whereas the Co(N(CN)$_{2}$)$_{2}$ undergoes a 
transition from ferromagnetic to antiferromagnetic interactions at nominally 13~kbar.\cite{pressure-dependence-Mdca} 
Herein, low and high field dc and ac magnetization studies for Mn(N(CN)$_{2}$)$_{2}$ are reported as a function of 
pressure up to 12.1~kbar. The data indicate an increase in the strength of the superexchange interaction with pressure 
and the appearance of a large magnetic anisotropy above 8.6~kbar. These results allow a $(P,T,H)$ phase diagram 
for Mn(N(CN)$_{2}$)$_{2}$ to be constructed. Finally, the study of a model Hamiltonian for 
this system suggests the pressure-induced changes in the spin-flop field and in the ordering temperature are 
driven by a change in the exchange anisotropy.

\section{\label{sec:level2}Experimental section}

To synthesize the Mn(N(CN)$_{2}$)$_{2}$ crystalline powder, a procedure described in the literature was followed.\cite{Batten1999} 
Specifically, Mn(ClO$_4$)$_2\cdot $6H$_2$O (1.81~g, 5~mmol) was mixed with Na(N(CN)$_{2}$)$_{2}$ (0.89~g, 10~mmol) and 2~mL of 
deionized water was added to the mixture. The solution was then heated to boiling for 10~minutes. The obtained white solid was 
washed with ethanol and diethyl ether. CHN analyses for MnN$_6$C$_4$: Calculated ($\%$): C,~25.69; H,~0.0; N,~44.92; 
Found ($\%$): C,~25.78; H,~0.0; N,~43.94. The FTIR absorption peaks in the region 2360~cm$^{-1}$ to 2192~cm$^{-1}$ are 
consistent with the tridentate binding mode of the dicyanamide ligand through the nitrile and amide N atoms.\cite{DivyaBook} 
In addition, the powder XRD peaks agree with the reported crystal structure of the title compound.\cite{Batten1999} 
The FTIR and XRD data sets are given as Figs.~\ref{fig:SM1} and \ref{fig:SM2} in the Supplemental Material (SM) at the end of 
the manuscript.

Using commercial Quantum Design MPMS-XL7 and MPMS-5S SQUID magnetometers, dc and ac magnetic measurements of as-grown 
crystalline powder of Mn(N(CN)$_{2}$)$_{2}$ were performed by employing standard techniques for the ambient pressure studies 
and a home-made pressure cell for the high pressure investigations. Specifically, for the ambient pressure studies, the sample was 
weighed $(\approx 12$~mg) and placed between two gelatin capsules, which were housed in a transparent drinking straw that was attached 
to a standard probe.  Contrastingly, the pressure cell, which is a modified self-clamping device,\cite{Thompson_pressurecell} is made  of
beryllium copper, the sample holder is made of teflon, and the pressure transmitting fluid is Daphne oil 7373.
Pressurization is achieved by the use of two screws that cap the ends of the cell body, while the superconducting transition temperature 
of Pb was used to determine the pressure at low temperatures and nominally $4$~mg of sample were loaded in the teflon can.\cite{PeprahThesis}

The sample was initially cooled and the magnetization measured from 6~K to 8~K in a field of 10~Oe to establish the superconducting 
transition temperature of the Pb. Then the sample was warmed to room temperature and field-cooled (fc) to a base temperature of 8~K 
in a field of 100~Oe, and the data were collected while warming. Next, the magnetic field was zeroed by using a degaussing 
sequence\cite{PajerowskiPhD} while at room temperature, and the sample was then cooled to a base temperature where a 100~Oe field was 
applied so the zero-field cooled (zfc) magnetization data could be collected while warming. Finally, the isothermal magnetization as a 
function of field was acquired at 8~K after field-cooling in a field of 100~Oe. The previous sequence was repeated for all the pressures 
studied. Even though the pressure was measured at the beginning of the sequence, before the temperature sweeps, additional studies show 
the pressure value at low temperatures is robust upon temperature and field cycling.\cite{QuinteroThesis} Upon release of the pressure, 
the magnetization values returned to the ones measured at ambient pressure, indicating the pressure-induced changes were 
completely reversible. The magnetic background signal of the beryllium copper pressure cell at low temperature was 
typically two orders of magnitude lower than the signal of the Mn(N(CN)$_{2}$)$_{2}$ samples being studied.

\section{\label{sec:level3}Results and discussion}

Prior to the pressure-dependent magnetization studies, the sample of Mn(N(CN)$_{2}$)$_{2}$ was measured between two gelatin capsules. 
The canted-antiferromagnetic ordering was observed at $T_{\mathrm c}=16.0$~K, where as the remanent magnetization and coercive field at 
$T=2$~K were $M_{\mathrm r}$~=~55~emu~Oe/mol and $H_{\mathrm c}=700$~Oe, in agreement with previously reported 
values.\cite{Kmety2000,Lappas2003,Spin-Flop-Mndca} Similar values were measured in the pressure cell near ambient pressure. 

\begin{figure}[b]
\includegraphics[width=\linewidth]{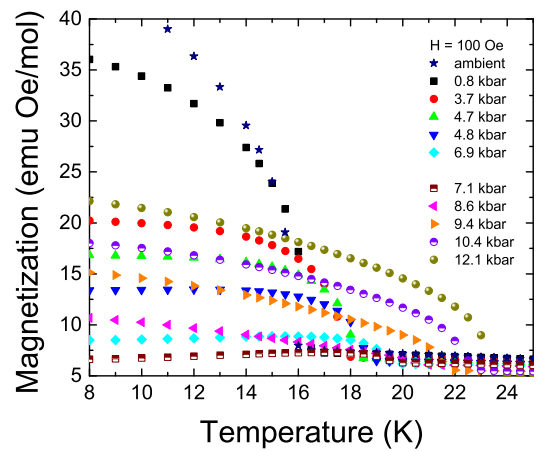}
\caption{\label{fig:FCdata}  (Color online)  Isobaric field-cooled (fc) magnetizations as a function of temperature and at 
the pressures given in the legend.  The cooling and measuring fields are both 100~Oe.}
\end{figure}

\subsection{\label{subsec:level1}Low magnetic field behavior}

The fc magnetization for 11 different pressures is shown in Fig.~\ref{fig:FCdata}. The value of the magnetization at 8~K 
decreases with increasing pressure for $ P < 7.1$~kbar, and the behavior is inverted for $ P > 7.1$~kbar, reaching a value 
of 22.1~emu~Oe/mol at the maximum pressure, which is half of the initial value at ambient pressure. The shape of the fc 
magnetization curves is also notably different above and below $7.1$~kbar, suggesting different magnetic anisotropy regimes. 
Specifically, for 3.7~kbar $< P <$ 6.9~kbar, the magnetization increases with decreasing temperature and quickly becomes 
temperature independent as expected for a system with low magnetic anisotropy.\cite{FC-ZFC-magnetization-1998} The data 
for $P=7.1$~kbar shows a large increase at $T_c$ and then constantly decreases, simulating the typical shape of a 
(non-canted) antiferromagnet. For $P>7.1$~kbar, the magnetization keeps increasing without reaching a plateau, and 
this behavior is associated with high anisotropy.\cite{FC-ZFC-magnetization-1998} At this point, it is important to clarify 
that a qualitative distinction between low and high magnetic anisotropy will be used during this subsection, but the 
high field data of the next subsection will allow an estimate the high magnetic anisotropy in this system.

\begin{figure}[b]
\includegraphics[width=\linewidth]{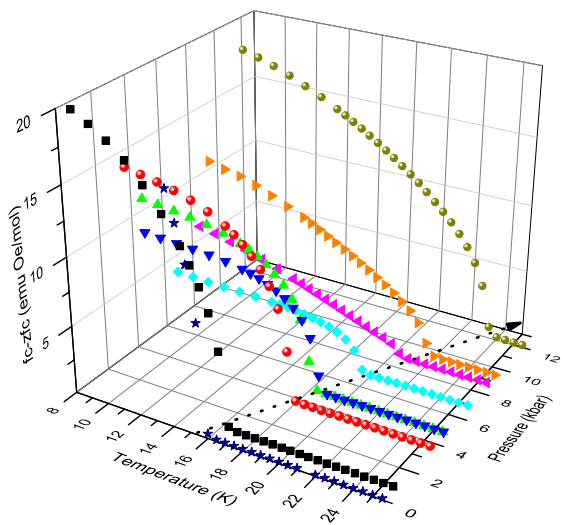}
\caption{\label{fig:3Dfc-zfc} (Color online) Isobaric fc-zfc magnetizations as a function of temperature are shown at 
several pressures.  The symbols and colors designating each pressure are the same as those used in Fig.~\ref{fig:FCdata}. 
For $P<~7.1$~kbar 
the fc-zfc magnetization decreases with pressure and the behavior is inverted for larger pressures. The dotted line serves 
as a guide for the eyes and represents the trend of $T_{\mathrm c}(P)$.  A detailed $T_{\mathrm c}(P)$ plot is given as 
Fig.~\ref{fig:TcvsP}.}
\end{figure}

The pressure-dependences of the differences between the fc and zfc magnetizations (fc-zfc) are shown in
Fig.~\ref{fig:3Dfc-zfc}. For $P<7.1$~kbar, the magnetization below the transition temperature  decreases with increasing pressure, 
and for $P>7.1$~kbar, the trend is inverted. This behavior derives from the fc magnetization given that the value of the 
zfc magnetization decreases monotonically with pressure, and the detailed data sets are presented in Fig.~\ref{fig:SM3} in the SM. 
The non-zero \mbox{fc-zfc} magnetization in this system can be attributed to spin-glass-like behavior or  
significant magnetic anisotropy. The shapes of the fc and zfc magnetization, where the signals quickly increase 
after $T_{\mathrm c}$ and then become flat or slowly increase, suggest the differences comes from magnetic anisotropy 
rather than glassy behavior.\cite{FC-ZFC-magnetization-1998,irreversible-SrRuO3} The lack of glassy behavior was confirmed 
by ac magnetic measurements at three different pressures and at different frequencies. For all pressures, the real component 
of the ac magnetization showed a transition temperature coincident with the value obtained from the dc measurement, and no 
frequency dependence of the magnetization was observed, Fig.~\ref{fig:SM4}.

The magnetic anisotropy plays a fundamental role in the differences between the fc and zfc magnetizations. During the 
zfc measurement, no external field is present when the sample is cooled through the ordering temperature, and the 
magnetic domains formed during the phase transition will have random orientations. Consequently, at base temperature 
when a small magnetic field of 100~Oe is applied, the magnetic response will depend on the magnitude of the anisotropy. 
For a low anisotropy system, the small field will be enough to reorient the domains in the direction of the field, and 
the magnetization will be similar to the response reflected in the fc data, making the fc-zfc magnetization small. 
By the same argument, a system with high anisotropy will show a large fc-zfc magnetization.

\begin{figure}[b]
\includegraphics[width=\linewidth]{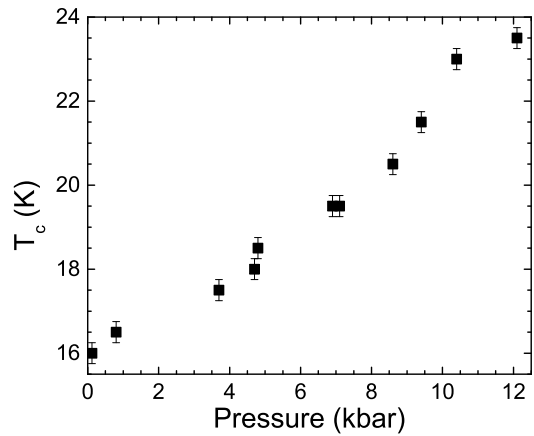}
\caption{\label{fig:TcvsP}  Transition temperatures of the long-range canted-antiferromagnetic order of 
Mn(N(CN)$_{2}$)$_{2}$ as a function of pressure, extracted from the fc data, Fig.~\ref{fig:FCdata}.}
\end{figure}

The temperature at which the canted-antiferromagnetic order occurs, $T_{\mathrm c}(P)$, increases with pressure over 
the entire range of pressures studied as shown in Fig.~\ref{fig:TcvsP}. The transition temperature increases from 
\mbox{$T_{\mathrm c}(P=\text{ambient}) = 16$~K} to \mbox{$T_{\mathrm c}$($P=12.1$~kbar) = 23.5~K}, 
see Fig.~\ref{fig:TcvsP}. This value corresponds to a change in $T_{\mathrm c}$ of 48$\%$  at 12.1~kbar, which is larger 
than the changes reported for the isostructural compounds M(N(CN)$_{2}$)$_{2}$ with  M = Fe, Co, and Ni, which show 
variations of up to 26$\%$ for the Ni analogue at 17~kbar.\cite{pressure-dependence-Mdca}

The pressure-induced enhancement of $T_{\mathrm c}$ can be understood in terms of an increase in the magnitude of 
the superexchange parameter $J$. The coupling of the metal ions in Mn(N(CN)$_{2}$)$_{2}$ is antiferromagnetic, and the 
Pauli principle suggest that the anti-parallel coupling between spins comes from the overlap of like orbitals (instead of 
unlike orbitals for ferromagnetic interaction).\cite{Goodenough} The overlap, then, increases with external pressure and,  
consequently, the magnetic interaction and transition temperature also increase.

The low field magnetization measurements suggest the following picture for Mn(N(CN)$_{2}$)$_{2}$. The pressure 
monotonically increases the strength of the antiferromagnetic interaction, $J$, favoring a smaller canting angle 
between the spins, thereby driving the fc and zfc magnetization to monotonically decrease with pressure while 
$T_{\mathrm c}(P)$ increases. However above 7.1~kbar, a large magnetocrystalline anisotropy appears, and as a result, 
a small field of 100~Oe is not enough to reorient the spins along the easy-axis, causing significant differences 
between the the fc and zfc data sets, Fig~\ref{fig:3Dfc-zfc}. Moreover the opposite pressure dependences of the 
fc and zfc magnetizations for $P>7.1$~kbar suggest the anisotropy is increasing with pressure. In the next subsection, 
the high field behavior of the magnetic response of Mn(N(CN)$_{2}$)$_{2}$ will be presented and discussed within the 
framework of this emerging interpretation. 

\subsection{\label{subsec:level2}High magnetic field behavior}

\begin{figure}[b]
\includegraphics[width=\linewidth]{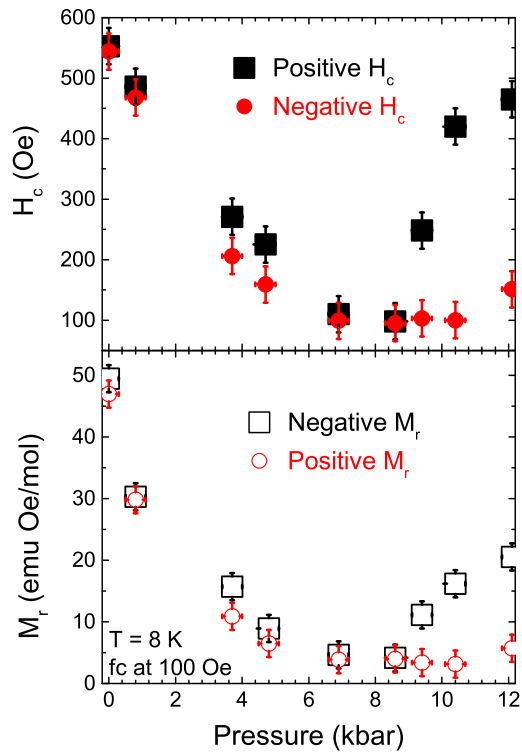}
\caption{\label{fig:MremandHc}  (Color online) Coercive fields and remanent magnetization values extracted from the field 
sweeps at $T =8$~K at different pressures, Fig.~\ref{fig:SM5}. The positive and negative coercive fields are defined as the crossing 
of the hysteresis loop with the positive and negative \mbox{x-axis}, and similarly, the positive and negative remanent 
magnetization are the crossings with the \mbox{y-axis}.}
\end{figure}

The field dependences of the fc magnetizations at $T=8$~K were measured at different pressures. 
Figure~\ref{fig:MremandHc} shows the coercive fields, $H_{\mathrm c}(P)$, and remanent magnetization values, 
$M_{\mathrm r}(P)$, extracted from each hysteresis loop. The complete hysteresis data sets are plotted in Fig.~\ref{fig:SM5}  
in the SM. The positive and negative $ H_{\mathrm c}(P)$ are defined as the crossing of the magnetic hysteresis loop 
with the positive and negative \mbox{x-axis}, respectively. In the same way, the positive and negative 
$M_{\mathrm r}(P)$ are defined as the crossing of the hysteresis loop with the positive and negative \mbox{y-axis}.

The coercive field decreases with increasing pressure for \mbox{$P<8.6$~kbar}, while the trend is inverted for 
\mbox{$P>8.6$~kbar}, and the same behavior is followed by the remanent magnetization values, as shown in 
Fig.~\ref{fig:MremandHc}. Additionally, for $P \leq 8.6$~kbar, the positive and negative coercive fields and 
remanent magnetization values are the same within experimental resolution, but for $P>$~8.6~kbar, they become 
visibly different. The difference between the positive and negative values for $ H_{\mathrm c}(P)$ and 
$M_{\mathrm r}(P)$ increases with pressure, reaching values of 314~Oe and 14.8~emu~Oe/mol, respectively, at 12.1~kbar.

The magnetic field necessary to flip a spin will increase if the magnetic anisotropy barrier increases. The  similarity 
of the pressure-dependent behavior of the coercivity and the low field \mbox{fc-zfc} magnetization, Fig.~\ref{fig:3Dfc-zfc}, 
is another signature that a change in the magnetic anisotropy is driving the behavior of  Mn(N(CN)$_{2}$)$_{2}$ for large pressures.

\begin{figure}[b]
\includegraphics[width=\linewidth]{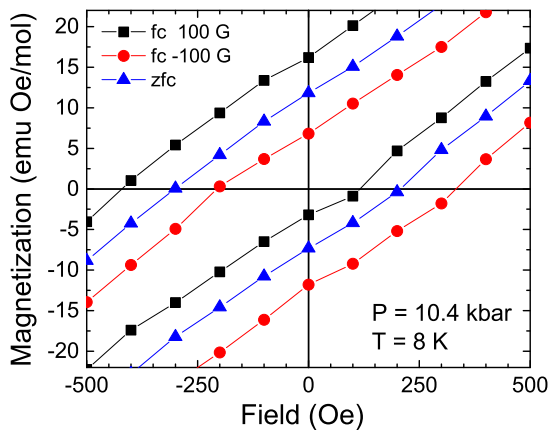}
\caption{\label{fig:1GPAfieldseeps} (Color online) An expanded view of the low-field regions of the experimentally accessible 
($\pm 70$~kOe) hysteresis loops after cooling in the fields indicated in the legend. The lines are a guide for the eye. 
The asymmetries are minor loop effects, due to the fact that the maximum applied field $ H_{\mathrm{max}}=70$~kOe is lower 
than the saturation field $ H_{\mathrm S}=304$~kOe.\cite{Brinzari-saturation} The existence of the minor loop effects at 
high pressures is a fingerprint of the large pressure-induced magnetic anisotropy in Mn(N(CN)$_{2}$)$_{2}$, see text.}
\end{figure}

To study the asymmetry of the magnetic hysteresis loops in more detail, the data collection sequence was repeated after 
cooling the sample from room temperature in different fields. For pressures lower than 8.6~kbar, the hysteresis curves 
were independent of the value and orientation of the cooling field, but this behavior changed at larger pressures. 
For example, a typical data set is shown in Fig.~\ref{fig:1GPAfieldseeps} for a pressure of 10.4~kbar. When the sample 
was cooled in 100~Oe, the hysteresis loop appeared shifted towards negative fields and positive magnetization, 
while the opposite behavior was observed when the cooling field was $-100$~Oe.  On the other hand, when cooled 
in zero field, the hysteresis loops appears roughly symmetric with respect to the origin. Even though the 
field-dependent shifts along the x-axis are consistent with what would be expected from an exchange-bias system, 
in the case of Mn(N(CN)$_{2}$)$_{2}$, these shifts are caused by an anisotropy-driven minor loop effect. 
Exchange-bias effects are ruled out since the maximum field used in our measurements, $H_{\mathrm{max}}(P) =70$~kOe 
is lower than the saturation field of Mn(N(CN)$_{2}$)$_{2}$, 
which has been previously measured to be $H_{\mathrm{sat}}=304$~kOe at 4~K.\cite{Brinzari-saturation} 
Moreover, the interactions between the metal centers are  antiferromagnetic in the range of pressure studied, 
and additionally the shifts along the y-axis are not expected in a typical exchange-bias system.\cite{EB-review1,EB-review2}

The reason for the minor loop effects above 8.6~kbar is the appearance of a large magnetic anisotropy that is not present 
at lower pressures, and the fact that the maximum applied field of 70~kOe is not enough to saturate the sample, 
\mbox{$ H_{\mathrm{max}}<H_{\mathrm{sat}}$}.\cite{EB-review1,EB-review2,comment-EBCoFenanoparticles} Furthermore, in the 
literature,\cite{EB-review2,comment-EBCoFenanoparticles,HmaxHArelation} a stronger bound is used, and minor loop effects 
are expected just when the maximum applied field is not enough to overcome the anisotropy of the system. According to this 
statement, the minor loops are present if \mbox{$ H_{\mathrm{max}}<H_{\mathrm A}$}, where $H_{\mathrm A}$ is the anisotropy 
field of the system. This phenomenological relationship suggests the anisotropy field is of the order of 70~kOe for 
Mn(N(CN)$_{2}$)$_{2}$ at pressures larger than 8.6~kbar. 

The explanation of the minor loop effect for a system with large magnetic anisotropy is as follows. When the sample is fc 
in a positive field through the ordering temperature, the domains are oriented in the direction of the field, and 
given the large magnetic anisotropy in the system, it will be hard to rotate the spins in a different direction. 
In particular, when the sample is at base temperature of 8~K, the maximum negative field applied of 
$H_{\mathrm{max}}=-70$~kOe is not enough to overcome the anisotropy and align the domains in the negative direction. 
Therefore, the magnetic field required to flip the spins from the negative to the positive direction is lower than 
than the field required to flip the spins in the opposite way, and as a result, positive $H_{\mathrm c}$ is lower than 
negative  $H_{\mathrm c}$. Naturally, for the fc protocol in a negative field, the behavior is inverted, and when zfc 
is used, the hysteresis loop is roughly symmetric. In other words, the high magnetic anisotropy is the reason why the 
system remembers the sign of the field used to cool through the ordering temperature.

\begin{figure}[b]
\includegraphics[width=\linewidth]{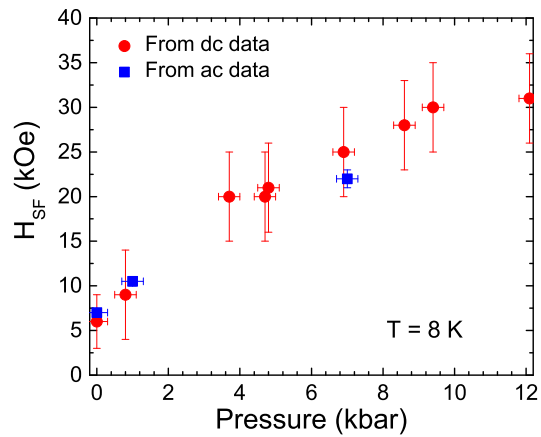}
\caption{\label{fig:spinflop} (Color online) Spin-flop fields for different pressures. The closed circle (red) data points 
are extracted from the location of the peak in the derivative of the dc magnetic hysteresis loops, and the closed square (blue) 
data points from the location of the peak in the real component of the ac magnetization, see Figs.~\ref{fig:SM6} and \ref{fig:SM7}.}
\end{figure}

\subsection{\label{subsec:leve3}Discussion}

Given the previously observed magnetoelastic coupling  in the M(N(CN)$_{2}$)$_{2}$ family,\cite{Lappas2003,Brinzari-saturation} 
the pressure-induced changes seen in Mn(N(CN)$_{2}$)$_{2}$ are most likely driven by magnetocrystalline anisotropy. 
Recent spectroscopic work at 300 K revealed a series of pressure-driven transitions in Mn(N(CN)$_{2}$)$_{2}$ with changes 
in the phonon behavior near $6$~kbar  and $17$~kbar. The transition at $6$~kbar was interpreted as a lattice distortion, 
while the more dramatic transition at $17$~kbar was associated with a reduction of the crystal symmetry.\cite{Brinzari-preprint} 
It is possible that the  pressure-induced magnetic anisotropy change seen in Mn(N(CN)$_{2}$)$_{2}$ at low temperature is driven 
by the same distortions of the lattice. However, crystallographic data as a function of pressure are necessary to confirm these conjectures.

\section{\label{sec:level4}Extended phase diagram}

\begin{figure}[t]
\includegraphics[width=\linewidth]{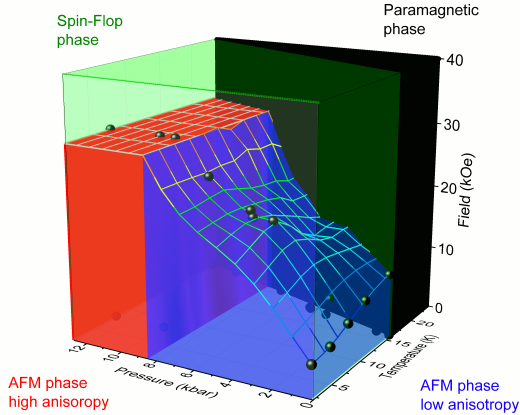}
\caption{\label{fig:pdwithpoints} (Color online) The $(P,T,H)$ phase diagram of Mn(N(CN)$_{2}$)$_{2}$. The black 
spheres are the data points from Figs.~\ref{fig:TcvsP} and \ref{fig:spinflop}, and those reported by Manson \emph{et al.}
\cite{Spin-Flop-Mndca} Four regions are identified by shading and correspond to the paramagnetic (PM) phase (black), 
the canted-antiferromagnetic (C-AFM) phase with low anisotropy (blue) and high anisotropy (red), and the spin-flop (SF) phase (green).}
\end{figure}

Using the high field magnetic data, the phase diagram of Mn(N(CN)$_{2}$)$_{2}$ can be explored. The spin-flop field 
for Mn(N(CN)$_{2}$)$_{2}$ can be observed using either ac or dc field-dependent magnetization 
measurements.\cite{Kmety2000,Lappas2003,Spin-Flop-Mndca} In the dc magnetic data, the spin-flop field appears as a 
peak in the derivative of the magnetization as a function of field, and in the ac magnetic data, as a peak in the 
in-phase component of the magnetization,  see Figs.~\ref{fig:SM6} and \ref{fig:SM7}. Figure~\ref{fig:spinflop} shows the pressure dependence 
of the the spin-flop field at base temperature extracted from the ac and dc field-dependent magnetization data. 
The spin-flop field of 7~kOe at ambient pressure and 8~K coincides with previous reports\cite{Spin-Flop-Mndca} and 
increases with pressure, reaching a value of 31~kOe at 12.1~kbar.

The $(P,T,H)$ phase diagram is shown in Fig.~\ref{fig:pdwithpoints}, where the data from Figs.~\ref{fig:TcvsP} and \ref{fig:spinflop} have been 
combined with the results of $H_{\mathrm{SF}}(T)$ reported by Manson \emph{et al.}\cite{Spin-Flop-Mndca}  Four regions 
have been identified by making some extrapolations of the existing data sets.  For example, the surface separating 
the AFM-PM regions is a horizontal wall, meaning that the transition temperature for all pressures is field independent 
up to 40 kOe.  In addition, the AFM-SF surface was extrapolated using the data at 8 K, and the separation from the 
low anisotropy and high anisotropy regions inside the AFM phase was marked at a field-independent pressure of 8.6 kbar.  
Finally, it is important to note that the magnetic field axis extends to 40 kOe, which is significantly lower than the 
saturation value of 304 kOe.\cite{Brinzari-saturation}

\section{\label{sec:level5}Model}

To develop a model for this compound in a magnetic field, the polycrystalline nature of sample is accommodated by 
averaging over all field orientations.  As a next step, consider a model with only (possibly anisotropic) 
nearest-neighbor interactions $J_{\alpha}$ between $S=5/2$ Mn$^{2+}$ spins.  In a magnetic field $H{\bf m}$ along 
${\bf m}$, the Hamiltonian ${\cal H}$ is then given by
\begin{equation}
{\cal H} = - \sum_{\langle i,j \rangle }J_{\alpha }S_{i\alpha }S_{j\alpha } - K \sum_i S_{ix}^2 - H \sum_i \bf{S}_i \cdot \bf{m}
\;\;\;,
\end{equation}
where $K$ is the easy-axis anisotropy that aligns the spins along the $x$ axis.   Due to the small canted 
moment of 0.002~$\mu_B$,\cite{Lappas2003} the small Dzyaloshinskii-Moriya interaction can be ignored.

Assuming that the exchange anisotropy is small, $J$ can be estimated from the saturation field in zero 
pressure.\cite{Brinzari-saturation}  Averaging over all field directions, $\bf m$, $H_{\mathrm{sat}} = 304$~kOe 
implies that $J = -0.087$~meV.  This exchange coupling then implies $T_{\mathrm{c}} = 23.5$~K, which overestimates 
$T_{\mathrm{c}}$ by about 50\%, as expected from mean-field theory. 

Now consider the origin of the spin-flop field, $H_{\mathrm{SF}}$, and its increase by a factor of 6 from 5~kOe at 
ambient pressure to 30~kOe at 12.1~kbar, Fig.~\ref{fig:spinflop}.  There are two possible origins for the spin-flop field. 
Firstly, $H_{\mathrm{SF}}$ may be caused by the easy-axis anisotropy $K$.  Such an anisotropy would be unexpected 
for $S=5/2$ Mn$^{2+}$ spins because its orbital angular momentum is quenched.  Nevertheless, after averaging over 
orientations of the magnetic field, one obtains
\begin{equation}
2\mu_{\mathrm{B}} H_{\mathrm{SF}} = 6.1 S \sqrt{|J K|} \;\;\;.
\end{equation}
Since $T_{\mathrm{c}} \propto \vert  J\vert$, the increase in $T_{\mathrm{c}}$ with pressure from 16~K to 24~K, 
Fig.~\ref{fig:TcvsP}, implies $\vert J\vert $ increases by about 50\%.  So the observed rise of $H_{\mathrm{SF}}$ 
from 5~kOe to 30~kOe requires that $K$ increases from $2 \times10^{-4}$~meV to $5 \times 10^{-2}$~meV, or an increase 
by a factor of 24.  This dramatic rise might occur due to a spin transition from $S=5/2$ to $S=3/2$ 
(an $S=1/2$ spin would also not have easy-axis anisotropy).  But there are two problems with this explanation.  
Firstly, a change in crystal field would result in both $e_g$ electrons 
paring with $t_{2g}$ electrons of the opposite spin, thereby producing $S=1/2$ not $S=3/2$.  Secondly, easy-axis anisotropy 
would cause $H_{\mathrm{SF}}(T)$ to drop with temperature from $H_{\mathrm{SF}}(T=0)$.\cite{PhysRevB.12.1908}    
However, $H_{\mathrm{SF}}(T)$ is observed to rise with temperature for this material, see Fig.~\ref{fig:pdwithpoints}.

The other possible origin for $H_{\mathrm{SF}}$ is anisotropic exchange $J_{\alpha }$ with $J_y = J_z \equiv J$ and 
$\Delta J = J_x - J_z < 0$, so the exchange favors antiferromagnetic alignment of the spins along the $x$ axis.  
Anisotropic exchange is believed to be present in many $S=5/2$ materials.\cite{Spin-Flop-Mndca}  In all such materials, 
$H_{\mathrm{SF}}(T)$ initially rises with temperature from its value at $T=0$, in agreement with the prediction by 
Rives and Benedict. \cite{PhysRevB.12.1908}  After averaging over orientations ${\bf m}$ of the field, one obtains
\begin{equation}
2\mu_{\mathrm{B}} H_{\mathrm{SF}} = 12.2 S \sqrt{ J \Delta J  }\;\;\;.
\end{equation}
At ambient pressure, this relation implies that $\Delta J = -4.2\times 10^{-5}$~meV so that the exchange anisotropy is 
$\Delta J /J= 4.8\times 10^{-4}$.

Since $T_{\mathrm{c}} \propto \vert J +\Delta J\vert $, the increase in $T_{\mathrm{c}}$ with pressure from 16~K to 24~K 
implies that $\vert J +\Delta J\vert $ rises by about 50\%.  The relations for $H_{\mathrm{SF}} $ and $T_{\mathrm{c}}$ imply 
that at high pressures, $J = -0.1296$ meV and $\Delta J = -1.0\times 10^{-3}$ meV.
So the exchange anisotropy $\Delta J/J $ rises from 0.05\% at ambient pressure to 0.8\% at 12.1 kbar. This last possibility 
seems like the most plausible explanation for the increase in $H_{\mathrm{SF}}$.   
Since $H_{\mathrm{sat}} $ depends very weakly on $\Delta J$, the saturation field should also rise by about 50\% with pressure, 
which will hopefully be verified by future measurements.

\section{\label{sec:level6}Conclusions}

The magnetic behavior of Mn(N(CN)$_{2}$)$_{2}$ was studied under hydrostatic pressure using dc and ac magnetometry. 
The long-range canted-antiferromagntic ordering temperature increases with pressure from 16~K at ambient pressure to 
23.5~K at 12~kbar, which corresponds to a change of 48$\%$, and this value is larger than those previously reported for 
the isostructural compounds M(N(CN)$_{2}$)$_{2}$ with M~=~Fe,~Co, and Ni. The \mbox{fc-zfc} magnetization, the coercive field, 
and the remanent magnetization values decrease as the applied pressure increases for $P<7.1$~kbar, and the behavior is inverted 
for $P>7.1$~kbar. Additionally, a field-cool dependent asymmetry in the magnetic hysteresis loop is observed at 8~K for $P>8.6$~kbar. 
All of these effects are understood in terms of a monotonic increase of the superexchange interaction with pressure and 
the appearance of an enhanced magnetic anisotropy. The spin-flop field was found to monotonically increase with pressure, 
and a phase diagram was sketched in temperature, magnetic field, and pressure space. Finally, the changes in the 
spin-flop field and the ordering temperature were shown to be consistent with an increase in the exchange anisotropy parameter 
$\Delta J/J $  from 0.05\% at ambient pressure to 0.8\% at 12.1 kbar.

\begin{acknowledgments}
This work is supported, in part, by NSF DMR-1202033
(MWM), DMR-1405439 (DRT), and DMR-1157490 
(NHMFL). Research by RSF is sponsored by the Office of Science, Materials Sciences and
Engineering Division, Office of Basis Energy Sciences, U.S. Department of
Energy. We gratefully acknowledge enlightening
conversations with Jamie L. Manson and Janice L. Musfeldt.
\end{acknowledgments}

\section*{Supplemental Material}  
Supplemental characterization data for Mn(N(CN)$_{2}$)$_{2}$ are
presented. Specifically, the XRD data, FTIR data, the dc zero-field-cooled
magnetization as a function of pressure, the ac magnetization as a
function of temperature and field for three different pressures, and
the isothermal dc magnetization as a function of field for different
pressures are shown.
\renewcommand{\thefigure}{S\arabic{figure}}
\setcounter{figure}{0}
\begin{figure}[h]
\includegraphics[width=\linewidth]{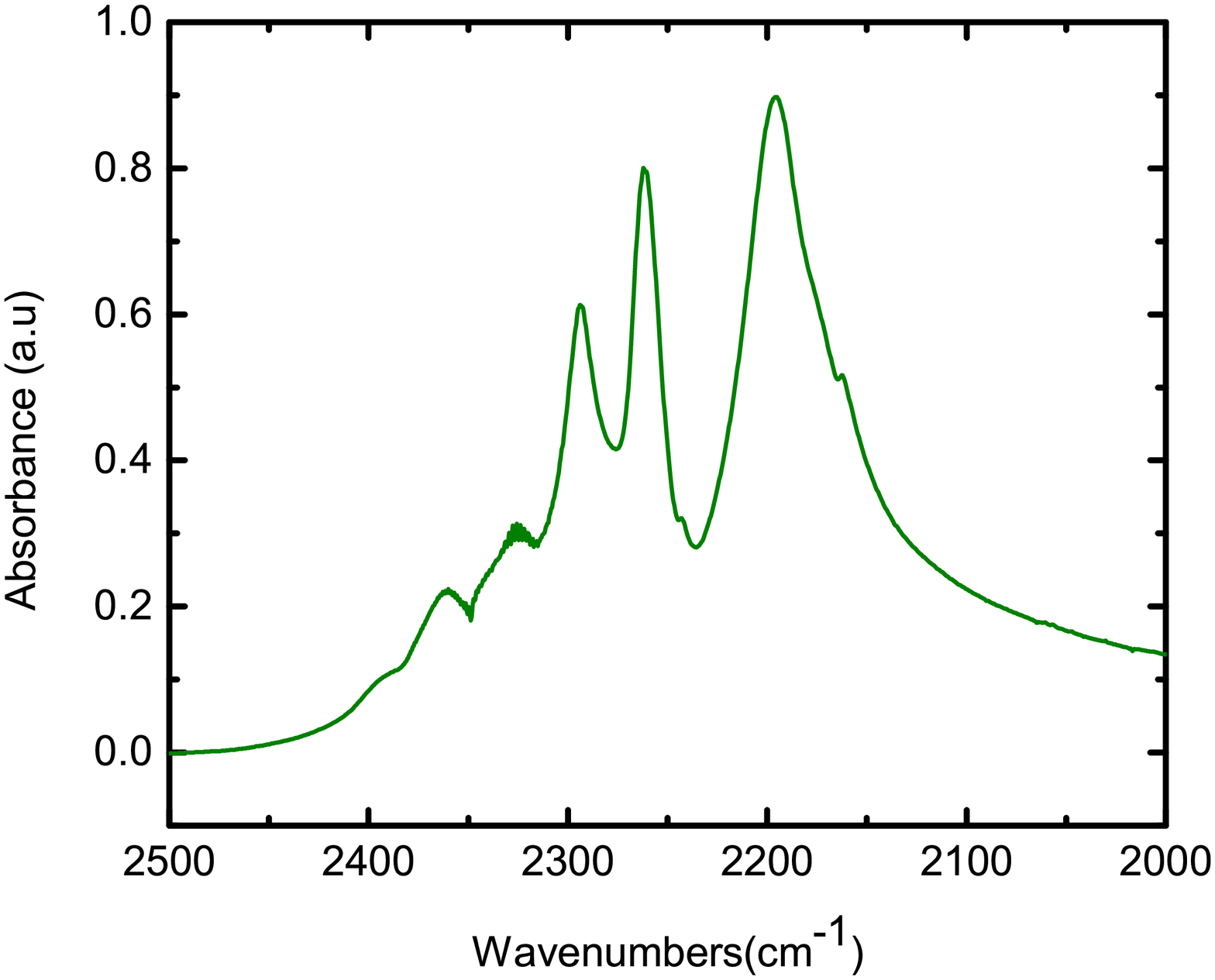}
\caption{\label{fig:SM1}FTIR spectra of Mn(N(CN)$_{2}$)$_{2}$. The peaks in the region 
2360~cm$^{-1}$ \textendash{} 2192~cm$^{-1}$ are consistent with the tridentate
binding mode of the dicyanamide ligand, through the nitrile and amide
N atoms (see main text).}
\end{figure}
\begin{figure}[h]
\includegraphics[width=\linewidth]{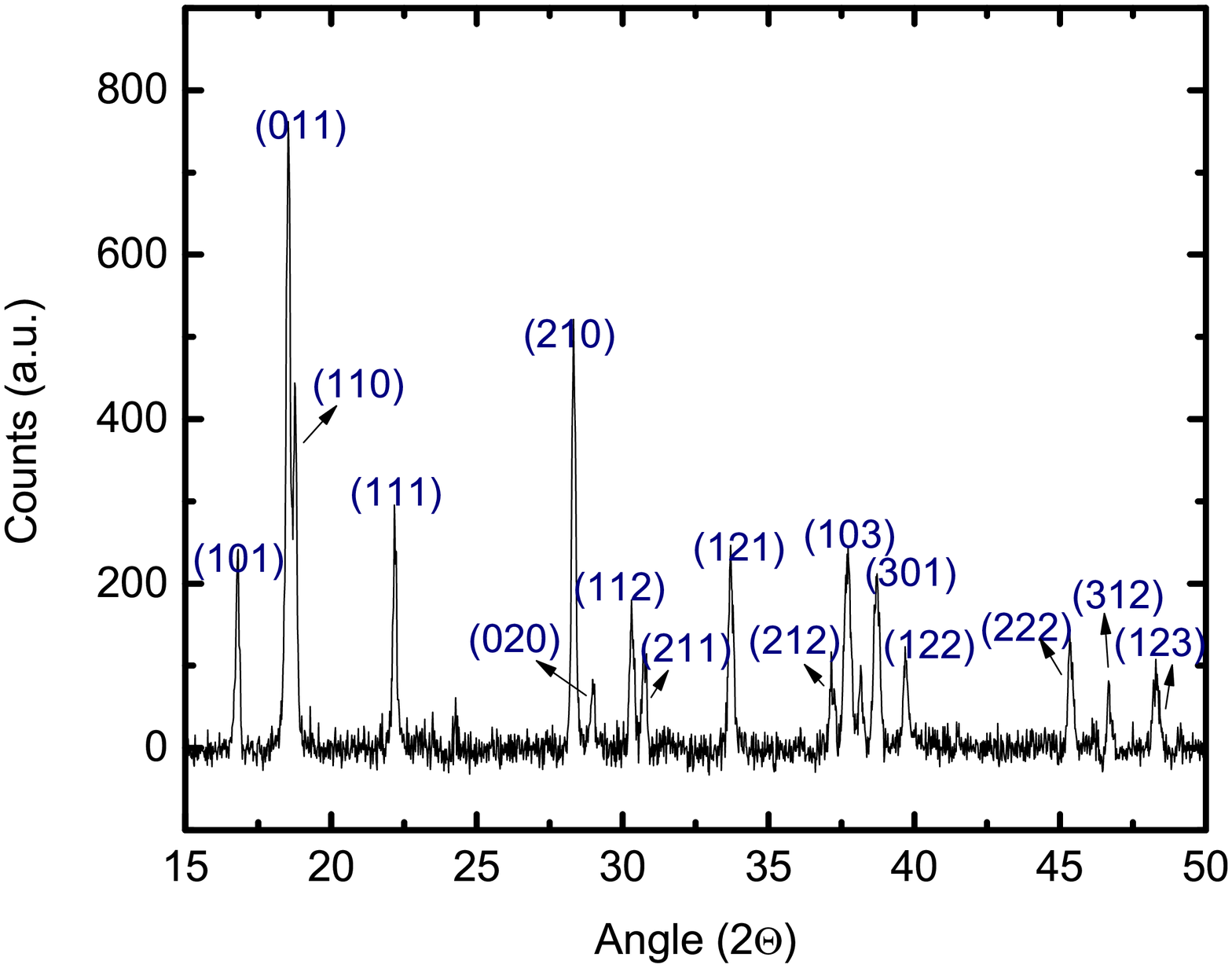}
\caption{\label{fig:SM2}Powder XRD spectra of Mn(N(CN)$_{2}$)$_{2}$. These data are consistent
with previously reported patterns (see main text).}
\end{figure}
\begin{figure}[h]
\includegraphics[width=\linewidth]{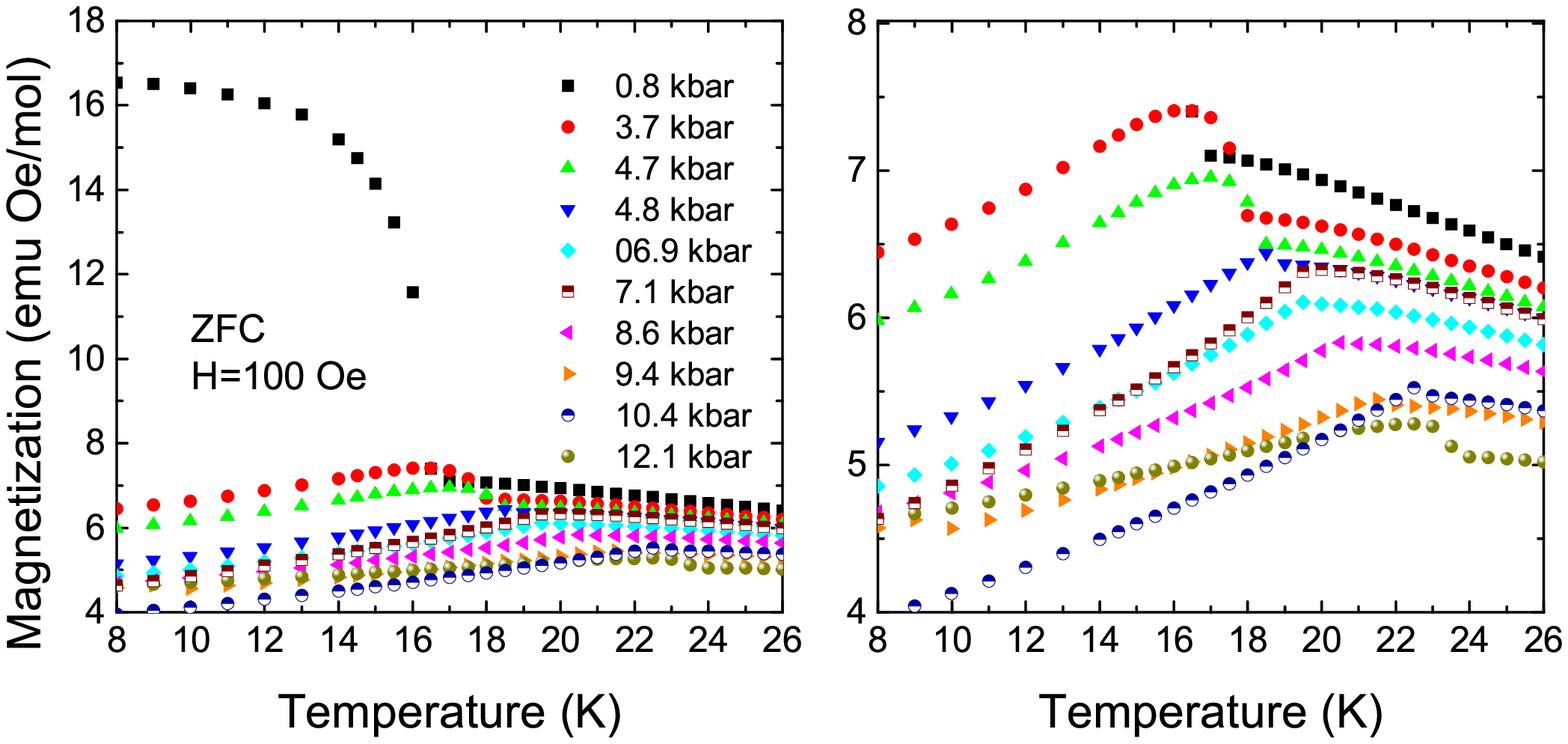}
\caption{\label{fig:SM3}Magnetization as a function of temperature for different pressures
measured in 100 Oe after zero field cooling from room temperature.
Both panels are the same set of data in different scales. The ordering
temperature increases with pressure while the magnetic response below
$T_{c}$ decreases with pressure (see main text).}
\end{figure}
\begin{figure}[h]
\includegraphics[width=\linewidth]{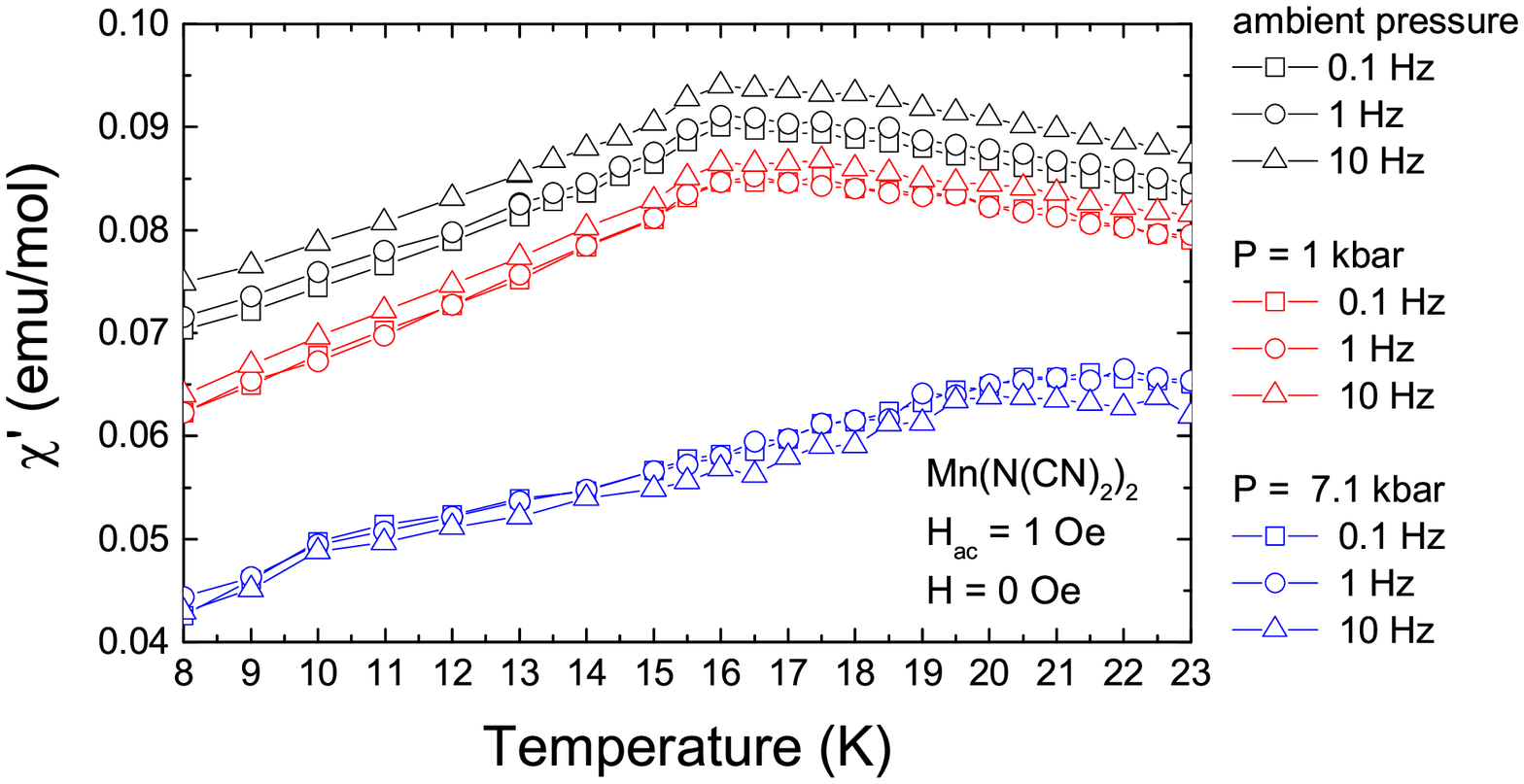}
\caption{\label{fig:SM4}In-phase component of the ac magnetization of Mn(N(CN)$_{2}$)$_{2}$
as a function of temperature for three different frequencies 0.1~Hz,
1~Hz, and 10~Hz at three different pressures, ambient, 1~kbar,
and 7.1~kbar. At frequencies larger than 10~Hz, scatter is present
due to the beryllium copper cell. The magnitude of the signal decreases
with pressure, in agreement with what was seen in the dc magnetic
data. At each pressure, a single peak is present at the ordering temperature,
and the data for different frequencies lie on top of each other within
the experimental error, suggesting the absence of spin-glass behavior
in the system (see main text).}
\end{figure}
\begin{figure}[h]
\includegraphics[width=\linewidth]{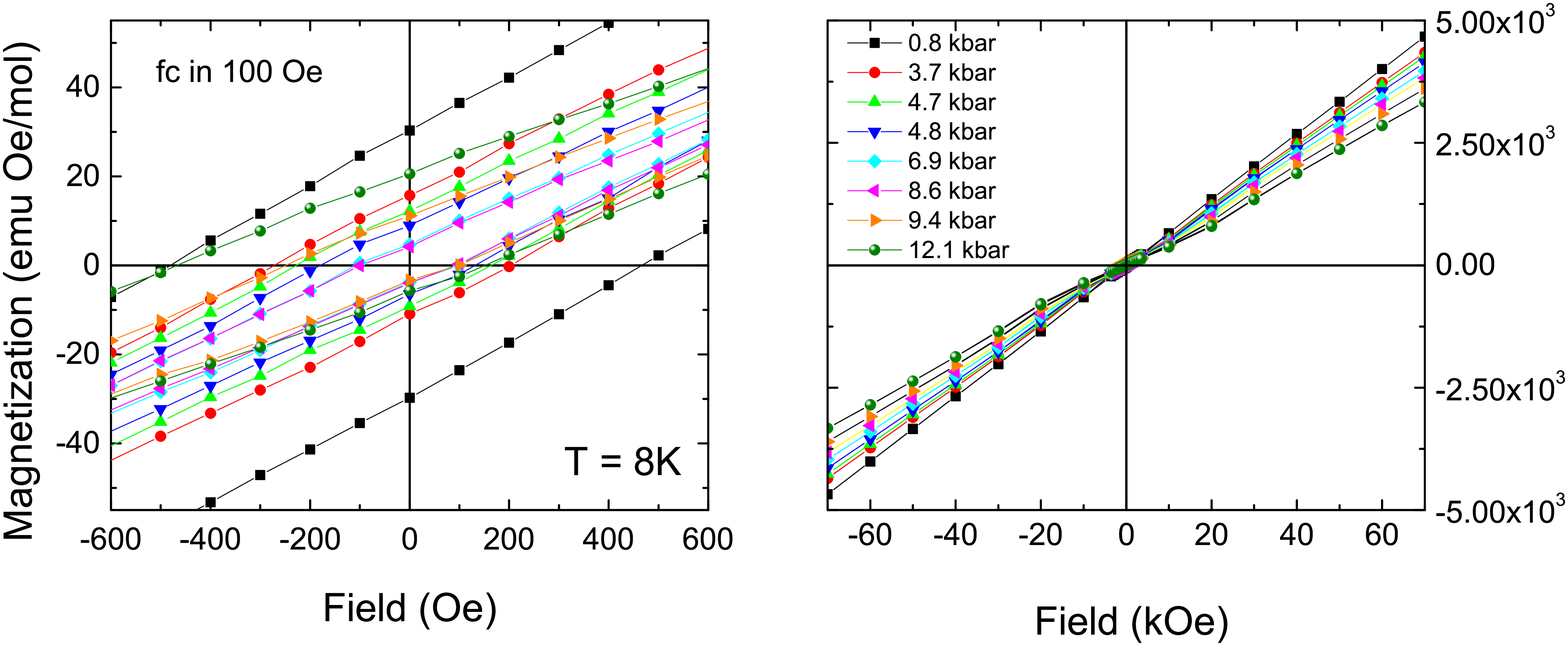}
\caption{\label{fig:SM5} The dc isothermal magnetization as a function of field at 8~K for different
pressures. Both panels contain the same set of data in different scales,
where the left panel shows the data at low fields and the right panel
shows the data in the entire range of fields $\pm70$~kOe. The coercive
field and remanent magnetization decrease with increasing pressure
for $P<8.6$~kbar, and increase with increasing pressures for $P>8.6$~kbar.
Additionally for $P>8.6$~kbar, the hysteresis loops become asymmetric,
and the asymmetry is particularly large for the data corresponding
to 12.1~kbar. The magnetization at 70~kOe decreases with increasing
pressure for all pressures. }
\end{figure}
\begin{figure}[h]
\includegraphics[width=\linewidth]{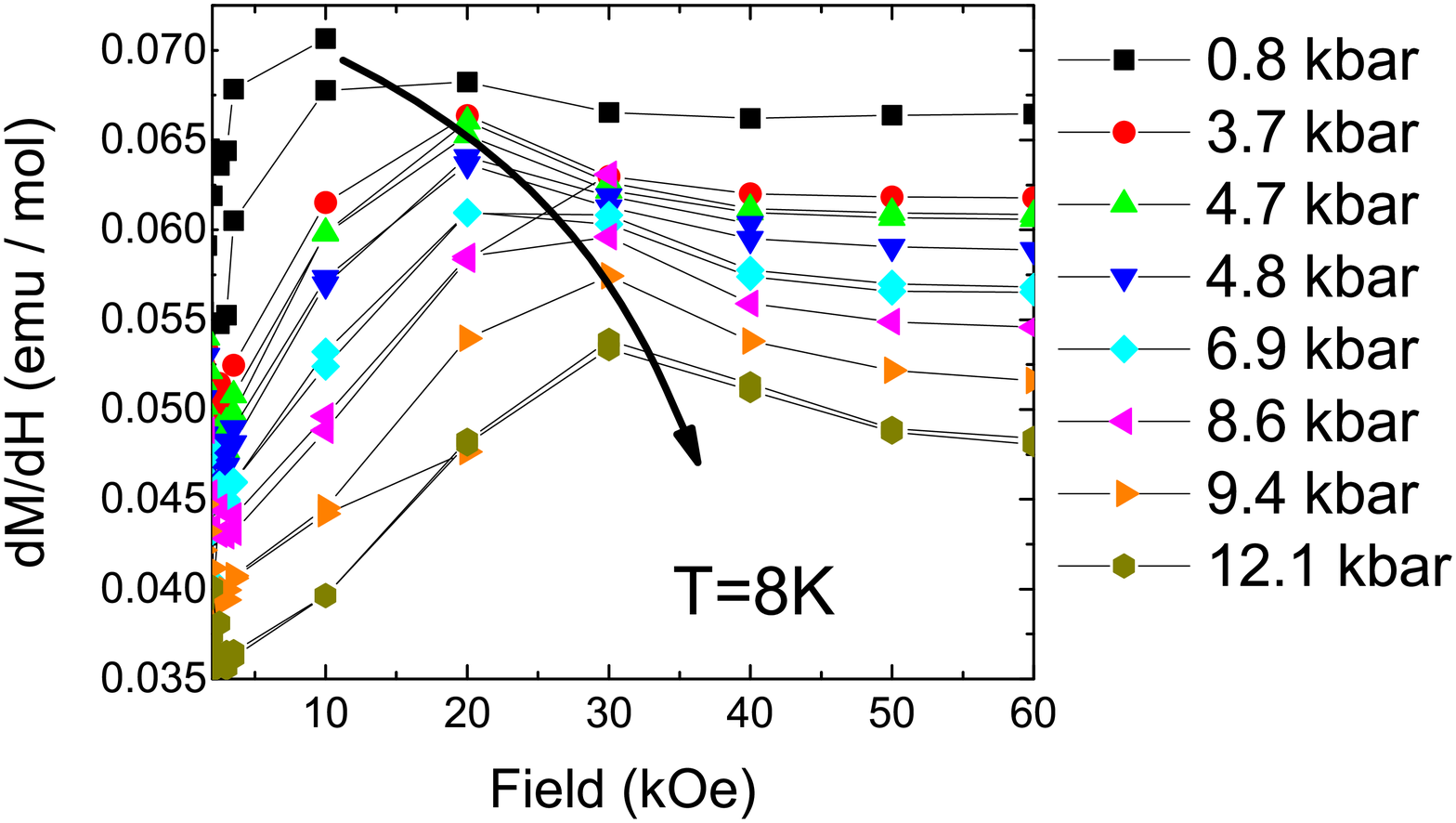}
\caption{\label{fig:SM6}Numerical derivative of the magnetization vs field
data for different pressures. The peak at each pressure corresponds
to the spin-flop field and increases with increasing pressure as indicated
by the black arrow (see main text).}
\end{figure}
\begin{figure}[h]
\includegraphics[width=\linewidth]{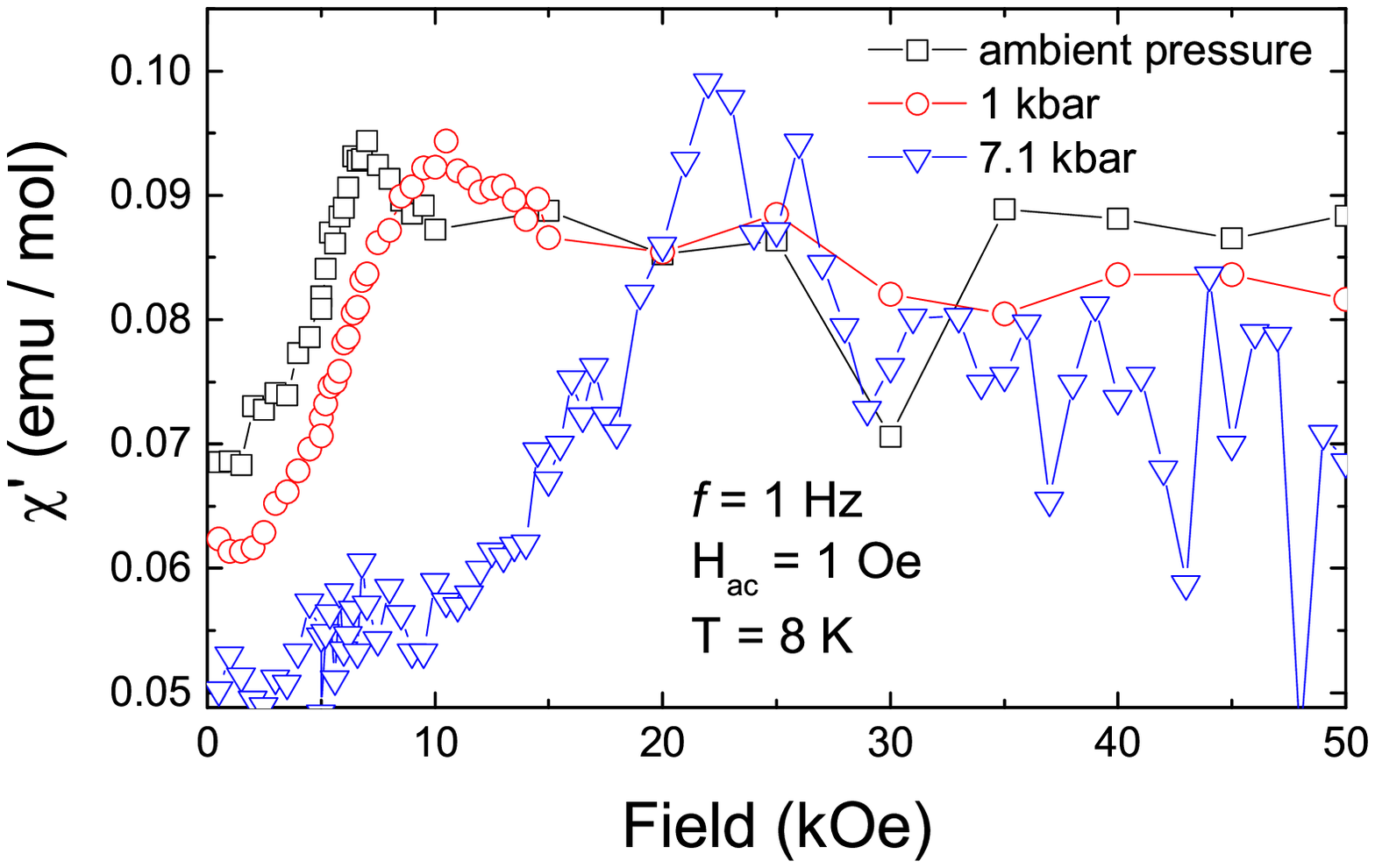}
\caption{\label{fig:SM7}In-phase ac magnetization as function of field for three different
pressures. The frequency was 1 kHz and the oscillating field had a
magnitude of 1~Oe. The peak corresponds to the spin-flop field, which
increases with pressure and confirms the behavior seen from the dc
data.}
\end{figure}

\clearpage

\bibliography{PAQ-Mndca2-arXiv}

\end{document}